\documentclass[12pt,epsf]{article}
\usepackage{epsfig}
\newcommand{\be}{\begin{eqnarray}}\newcommand{\ee}{\end{eqnarray}}
\newcommand{\beo}{\begin{eqnarray*}}\newcommand{\eeo}{\end{eqnarray*}}
\newcommand{\ba}{\begin{array}}\newcommand{\ea}{\end{array}}
\newcommand{\no}{\nonumber}

\begin{document}
\baselineskip=18pt
\parskip=3pt
\begin{titlepage}
\title{Non-orientable Boundary Superstring Field Theory\\with Tachyon Field}

\author{K. S. Viswanathan\footnote{kviswana@sfu.ca}\hspace{.2 cm}
and Yi Yang\footnote{yyangc@sfu.ca}}
\date{\textit{Department of Physics, Simon Fraser University\\Burnaby, BC V5A 1S6 Canada}}
\maketitle

\thispagestyle{empty}

\vspace{2.5 cm} We use the BSFT method to study the non-orientable
open string field theory (type I). The partition function on the M\"obius strip
is calculated. We find that, at the one-loop level, the divergence
coming from planar graph and unoriented graph cancel each other as
expected.

\vspace{5 cm} October, 2001
\end{titlepage}

\section{Introduction}

Tachyon condensation is an interesting issue in string theory
\cite{3,9911116,0010240}. To study it, Witten's Boundary String
Field Theory (BSFT) \cite{witten}-\cite{9303143} has proven to be
a useful method. Tachyon condensation at tree level has been
studied first by A. Gerasimov and S. Shatashvili \cite{0009103}
and D. Kutasov, M. Marino, G. Moore \cite{0009148}, then followed
by many authors \cite{0012089}-\cite{0105245}. The BSFT method was
also extended to the case of superstrings in \cite{0003101},
\cite{0010108}-\cite{0012198}. Recently, tachyon condensation at
one-loop level has been discussed by many authors who have studied
it by different methods and got similar results
\cite{0101207}-\cite{0106033}.

However, all of these considerations focus on type II superstring
theories which are oriented string theories, i.e. including only
annulus or cylinder diagrams at one-loop level. In this work, we
will study type I superstring theory. In fact, with Chan-Paton
charges, open-string theory in supersymmetric 10 dimensional
space-time (type I) has to be non-orientable \cite{book1}. To
study the non-orientable open-string theory, we need to include
the M\"obius strip at the one-loop level.

It is well known that the gauge group in type I superstring theory
has to be $SO(32)$. There are several ways to verify this by
anomaly or divergence cancellation. In \cite{book1,book2}, the
on-shell $n$ strings amplitudes for different one-loop graphs are
calculated. It was shown that the divergences, due to the
integration over the modulus parameter, coming from planar and
M\"obius graphs are cancelled each other only for $SO(32)$ gauge
group. By using BSFT, we will show that, off-shell, this
divergences cancellation also holds only for $SO(32)$ gauge group.

In the next section, we briefly recall the analysis of powers of
the coupling constant in non-orientable string graphs. In the
section 3, the partition function on the M\"obius strip is
calculated by the method of BSFT. In the section 4, we include
Chan-Paton factors by considering an unstable $Dp$-brane in $N$
$D9$-branes background, then show that the divergences due to the
integration over the modulus parameter only cancel each other for
$N=32$, namely $SO(32)$ gauge group. We summarize our work in the
section 5.

\section{Powers of the Coupling Constant in String Graphs}

Boundary string theory is defined on all compact two dimensional
manifolds \cite{witten}. At each order the string coupling
constant enters the calculation as:
\be
g^{-\chi}. \ee where $g$ is open string coupling and $\chi$ is the
Euler number of the manifold.

Any compact orientable two-manifold is topologically equivalent to
the direct sum of a sphere with $h$ handles and $b$ holes, and the
corresponding Euler number is
\be
\chi=2-2h-b. \ee

Similarly, any compact non-orientable two-manifold is
topologically equivalent to a direct sum of a sphere with $h$
handles, $b$ holes and $c$ cross-caps. The corresponding Euler
number is \cite{burgess}
\be
\chi=2-2h-b-c. \ee

It is well known that Type I superstring theory has to be a
unoriented open string theory. We will use the string coupling
power counting to decide what two-manifolds, which can couple to
the D-brane, we need.

\begin{figure}
\epsfxsize=11.9cm \centerline{\epsffile{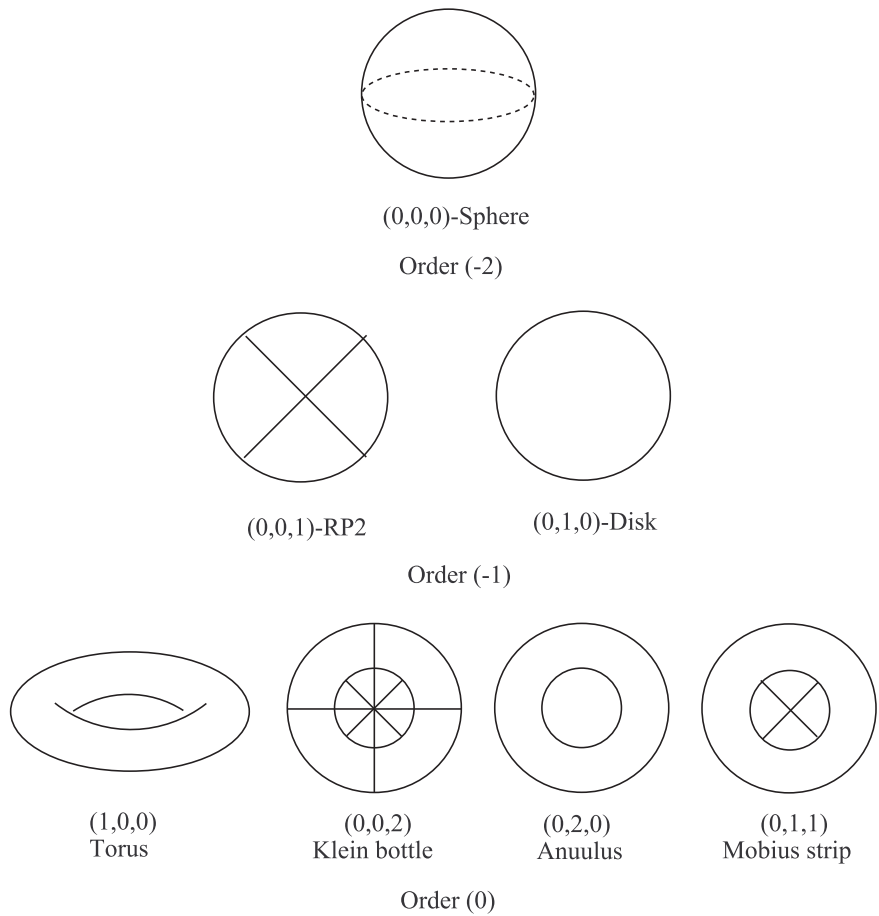}} \vspace{0cm}
\caption{compact non-orientable two-manifold} \label{powers}
\end{figure}

The lowest order ($g^{-2}$) is the sphere for which
$(h,b,c)=(0,0,0)$. We will not consider it because it can not
couple to the D-brane. At the next order ($g^{-1}$) there are two
graphs, the $RP_2$ with $(h,b,c)=(0,0,1)$ and the disc with
$(h,b,c)=(0,1,0)$. For the same reason as for the sphere, we will
only consider the disc. At the one-loop order ($g^{0}$) there are
four graphs, which are the torus with $(h,b,c)=(1,0,0)$, the
annulus with $(h,b,c)=(0,2,0)$, the Klein bottle with
$(h,b,c)=(0,0,2)$ and the M\"obius strip with $(h,b,c)=(0,1,1)$.
We will only consider the annulus and the M\"obius strip, the two
of them have boundaries. (Fig. \ref{powers})

The partition function on the disc has been studied by many
authors. We will concentrate on the one-loop graphs in this work.

For open string theory, there are three types of one-loop diagrams
\cite{book1}. The first type is the planar diagram, which is an
annulus with only one boundary coupled to the D-brane. The second
type is the unoriented diagram, which is a M\"obius strip. Third
type is the non-planar diagram, which is an annulus with both
boundaries coupled to the D-brane.

The first and the third types have been considered in
\cite{0104099} and other papers. In the next section, we will use
the same method as in \cite{0104099} to calculate the partition
function on the M\"obius strip, then write down the effective
tachyon action for the unoriented superstring.

\section{Partition Function on M\"obius Strip}

\begin{figure}
\epsfxsize=15cm \centerline{\epsffile{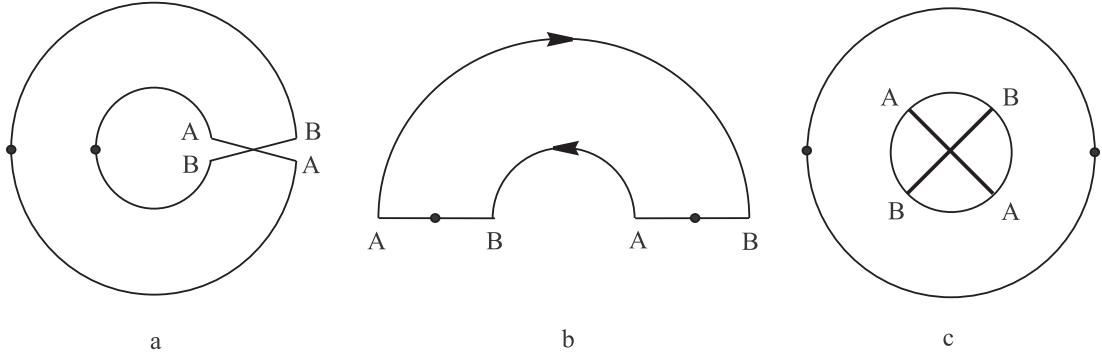}} \vspace{0cm}
\caption{Three descriptions of the M\"obius strip} \label{mobius}
\end{figure}

M\"obius strip can be described as an annulus with a twist (a of
Fig. \ref{mobius}). But it is difficult to give it global
coordinates. Another description is mapping the M\"obius strip to
the upper half of an annulus with the upper semicircle identified
with the lower in an anti-parallel way (b of Fig. \ref{mobius}).
This description has been widely used to calculate the Green
function on the M\"obius strip world sheet
\cite{book1,0003180,0105307}. But we will have problems when
trying to embed the M\"obius strip in spacetime and attach to the
D-brane. The third description is to use a "cross-cap" (c of Fig.
\ref{mobius}). In this description, the M\"obius strip looks like
an annulus, but one of the boundaries is replaced by a cross-cap.
We will use the cross-cap description to calculate the Green
function on the M\"obius strip in the following.

Now the world sheet $\Sigma$ is a M\"obius strip with a
rotationally invariant flat metric \be
ds^2=d\sigma_1^2+d\sigma_2^2, \ee the complex variable
$z=\sigma_1+i\sigma_2$, with $a\leq |z|\leq b$. Here we choose
$\rho=a$ as the boundary of the M\"obius strip, and $\rho=b$ is
the cross-cap.

Next, we take the tachyon profile as \be
T(X)=uX;\hspace{1cm}\rho=a,\ee the boundary term can then be
written as \be I_{\rm bndy}=\frac{y}{8\pi}\int_0^{2\pi}
d\theta\left(X^2+\psi\frac{1}{\partial_\theta}\psi
+\tilde{\psi}\frac{1}{\partial_\theta}\tilde{\psi}\right)_{\rho=a},
\ee where $y\equiv u^2$.

The boundary conditions for the Green function of the bosonic
fields are \cite{0104099}
\be
\label{bc} (z\partial +\bar z\bar\partial
-y_a)G_B(z,w)|_{\rho=a}&=&0,\no\\
G_B(z,w)|_{\rho=b}-G_B(-z,w)|_{\rho=b}&=&0, \ee where the
condition at the boundary $\rho=a$ is same as the condition for an
annulus, while at the boundary $\rho=b$ we have the symmetric
condition for the cross-cap.

To solve for $G_B(z,w)$, we start with the ansatz,
\be\label{ansatz}
G_B(z,w)&=&-\ln |z-w|^2+C_1\ln |z|^2\ln |w|^2+C_2(\ln |z|^2+\ln |w|^2)+C_3\no\\
& &+\sum_{-\infty}^{\infty}a_k[(z\bar w)^k+(\bar z w)^k]
+\sum_{-\infty}^{\infty}b_k\left[\left(\frac{z}{w}\right)^k
+\left(\frac{\bar z}{\bar w}\right)^k\right]
\ee

One can easily verify that these are solutions of \be
\partial_z\bar\partial_{\bar z}G(z,w)=-2\pi\delta^{(2)}(z-w),
\ee on a M\"obius strip. $C_1$, $C_2$, $C_3$, $a_r$'s and $b_r$'s
are coefficients to be determined by the boundary conditions
(\ref{bc}). In the appendix, we derive the Green function and find
that the only differences consist of replacing
$\left(\frac{a^2}{b^2}\right)$ by $\left(-\frac{a^2}{b^2}\right)$
and setting $y_b=0$.

The fermionic Green functions satisfy the equations\footnote{For
simplification, in this work, we only consider the NS-NS sector
with spin structures $G_{++}$ and $G_{--}$ (see \cite{0105238} for
detail). The other spin structures $G_{+-}$ and $G_{-+}$ do not
contribute to the zero mode which the divergence cancellation
depends on. Moreover, the R-R section does contribute to the zero
mode, but after we set $y_b=0$ (for planar graph), it does not
affect the cancellation.}: \be
\bar{\partial}G_F(z,w)&=&-i\sqrt{zw}\delta^{(2)} (z-w),\no\\
\partial\tilde{G}_F(\bar{z},\bar{w})&=&+i\sqrt{\bar{z}\bar{w}}\delta^{(2)} (\bar{z}-\bar{w})
\ee with the boundary conditions
\be
\left.\left(1-iy_a\frac{1}{\partial_\theta}\right)G_F\right|_{\rho=a}
&=&\left.\left(1+iy_a\frac{1}{\partial_\theta}\right)\tilde{G}_F\right|_{\rho=a},\no\\
G_F(z,w)|_{\rho=b}&=&\tilde{G}_F(-\bar z,\bar w)|_{\rho=b}. \ee
where, similarly, the condition at the boundary $\rho=a$ is same
as the condition for an annulus, the condition at the boundary
$\rho=b$ is the symmetric condition for the cross-cap.

As in the bosonic case, we find that the fermionic Green function
on a M\"obius strip is the same as that on an annulus by replacing
$\left(\frac{a^2}{b^2}\right)$ by $\left(-\frac{a^2}{b^2}\right)$
and setting $y_b=0$.

The partition function, then, can be obtained as (see Appendix for
the detailed derivation) \be
Z_M(a^2/b^2)=Z'4^y\frac{Z^2_B(y,-a^2/b^2)}{Z_B(2y,\sqrt{-a^2/b^2})},
\ee where $Z'$ is the integration constant, and \be
Z_B(y;a,b)=\frac{1}{\sqrt{y}}\cdot\prod_{k=1}^\infty\left[\left(1+\frac{y}{k}\right)
-\left(1-\frac{y}{k}\right)\left(-\frac{a^2}{b^2}\right)^{k}\right]^{-1}
\ee is the bosonic partition function on the M\"obius strip.

Comparing this result with that for the planar graph
\cite{0104099}, we have, up to a constant, \be
Z_M\left(y,\frac{a^2}{b^2}\right)=Z_P\left(y,-\frac{a^2}{b^2}\right).
\ee

It is convenient to take the outer radius $b=1$ in the following.
Integrating over the modulus $d\lambda=da/a$, we obtain
\be\label{par} Z_M(y)&=&\int d\lambda Z_M(y,a^2)\no\\
&=&\sqrt{2}Z'4^{y}\int_0^1\frac{da}{a}\frac{1}{\sqrt{y}}\cdot\prod_{k=1}^\infty\frac{(1+\frac{2y}{k})
-(1-\frac{2y}{k})(\sqrt{-a^2})^k}
{\left[(1+\frac{y}{k})-(1-\frac{y}{k}) (-a^2)^{k}\right]^2}. \ee

Expanding (\ref{par}) in powers of $y$, the expression of $Z_M$
takes the form \be
Z_M(y)&=&\sqrt{2}Z'\int_0^1\frac{da}{a}\prod_{k=1}^\infty\left(\frac{1-(\sqrt{-a^2})^k}{[{1-(-a^2)^k}]^2}\right)\no\\
& &\cdot\left(\sqrt{\frac{1}{y}}+\left[2\ln {2}
-4\sum_{n=1}^\infty\ln (1-(-\sqrt{a^2})^{2n-1})\right]
\sqrt{y}+\cdots\right). \ee

\section{Divergence Cancellation for Type I Superstring Theory}

In this section, we will consider the type I superstring theory.
To study the tachyon condensation in the type I theory with gauge
group $SO(N)$, we can have two kinds of unstable $D$-brane
systems:

\begin{itemize}
  \item An unstable\footnote{To be unstable in Type I thoey, $p\neq$ 1, 5 or
  9.} $Dp$-brane in $N$ stable $D9$-branes.
  \item $N_1$ $D9$-branes and $N_2$ $\bar D9$-branes with $N_1-N_2=N$.
\end{itemize}

By brane-antibrane creation and annihilation, these two cases are
equivalent to each other \cite{9810188,9812031}. It is enough to
consider one of them. We will focus on the first case in this
paper.

Type I superstring theory is a non-orientable string theory,
therefore instead of a $U(N)$ group, $N$ $D9$-branes lead to a
nontrivial Chan-Paton $SO(N)$ space in $d$ dimensional
space-time\footnote{Later we will set d=10 for Type I superstring
theory.}. It is easy to include Chan-Paton factors by taking the
boundary action to be \be \label{non-bndy}e^{-S_{\rm bndy}}={\rm
Tr} P \exp\left\{-\frac{1}{8\pi}\int
d\theta\left[T^2(X)+(\psi^{\mu}\partial_{\mu}T)
\frac{1}{\partial_{\theta}}(\psi^{\nu}\partial_{\nu}T)\right]\right\}
.\ee

We can generalize the tachyon profile to the form as in
\cite{0010108} \be\label{non-tachyon}
T(X)=\sum_{\alpha=1}^{2n}T^\alpha(X)\gamma^\alpha \ee where we set
$N=2^n$ and expand the $2^n\times 2^n$ matrix tachyon field in
real Hermitian matrices $\gamma^\alpha$'s, which form a Clifford
algebra. The theory has $SO(2^n)$ gauge invariance, of which a
$SO(2n)$ subgroup is manifest. \cite{0012198}

In the following, we will consider the linear tachyon profiles of
the forms \be\label{non-linear}
T^\alpha(X)=\sum_{i=0}^{p}u^\alpha_iX^i, \ee where $i=0,\cdots,p$
are indices of world-volume of $Dp$-brane, and
$\alpha=1,\cdots,2n$ are indices of spinor representation of
$SO(2n)$ group.

Substituting (\ref{non-tachyon}) and (\ref{non-linear}) into
(\ref{non-bndy}), we find (see Appendix A.3) \be e^{-S_{\rm
bndy}}=2^n \exp\left\{-\frac{y_{ij}}{8\pi}\int
d\theta\left[X^iX^j+\psi^i\frac{1}{\partial_{\theta}}\psi^j
+\tilde{\psi}^i\frac{1}{\partial_{\theta}}\tilde{\psi}^j\right]\right\}
,\ee where
$y_{ij}\equiv\sum_{\alpha=1}^{2n}u^{\alpha}_iu^{\alpha}_j$.

The partition function on the M\"obius strip will simply be \be
Z_M(y,a^2)=-2^n\prod_{i=0}^pZ'_MZ_M(y_i,a^2), \ee where $Z'_M$ is
a integration constant and\footnote{Because
$G^{ij}\sim\delta^{ij}$, the terms of $y_{ij}$ with $i\neq j$ does
not contribute to the partition function.} $y_i\equiv y_{ii}$. The
overall negative sign comes from the twist operator for $SO(2^n)$
group \cite{book1}.

Combining the unoriented graph and the planar graph, we
obtain\footnote{The $(2^{n})^2$ in front of the planar partition
function comes from the two boundaries of an annulus
\cite{book1}.} \be Z_{M+P}(y)
&=&(2^{n})^2\int_0^1\frac{da}{a}\prod_{i=0}^pZ'_PZ_P(y_i,a^2)
-2^n\int_0^1\frac{da}{a}\prod_{i=0}^pZ'_MZ_M(y_i,a^2)\no\\
&=&(2^{n})^2\int_0^1\frac{da^2}{2a^2}\prod_{i=0}^pZ'_PZ_P(y_i,a^2)
-2^n\int_0^1\frac{da^2}{2a^2}\prod_{i=0}^pZ'_MZ_M(y_i,a^2)\no\\
&=&(2^{n})^2\int_0^1\frac{da^2}{2a^2}\prod_{i=0}^pZ'_PZ_P(y_i,a^2)
-2^n\int_0^1\frac{da^2}{2a^2}\prod_{i=0}^pZ'_MZ_P(y_i,-a^2)\no\\
&=&(2^{n})^2\int_0^1\frac{da^2}{2a^2}\prod_{i=0}^pZ'_PZ_P(y_i,a^2)
+2^n\int_{-1}^0\frac{da^2}{2a^2}\prod_{i=0}^pZ'_MZ_P(y_i,a^2).\ee

To proceed, we need the exact relationship between the two
integration constants $Z'_P$ and $Z'_M$. This can be done by
taking the ratio of two partition functions in some limit and
compare them to the direct calculation. Taking the limit

\be \lim_{a\rightarrow0}\lim_{y_i\rightarrow0}
\frac{Z'_PZ_P(y_i,a^2)} {Z'_MZ_M(y_i,a^2)}=\frac{Z'_P}{Z'_M}. \ee

On the other hand

\be \lim_{a\rightarrow0}\lim_{y_i\rightarrow0}
\frac{Z'_PZ_P(y_i,a^2)} {Z'_MZ_M(y^{\alpha}_i,a^2)}=
\lim_{a\rightarrow0}\frac{\int dXe^{I^P_{\rm bulk}}}{\int
dXe^{I^M_{\rm
bulk}}}=\lim_{a\rightarrow0}\frac{Z^{(0)}_P}{Z^{(0)}_M},\ee where
${Z^{(0)}_P}$ and ${Z^{(0)}_M}$ are the non-tachyon-interaction
partition functions for planar and unoriented graphs.

Fortunately, the non-tachyon-interaction partition functions for
planar and unoriented graphs have been calculated, they are
\cite{book2}
\be
Z^{(0)}_P&=&\left(\frac{1}{8\pi(8\pi^2\alpha')^{\frac{d}{2}}}\right)^{\frac{1}{d}}
\int_0^{\infty}ds[2(d-2)+O(e^{-2s})]\no\\
Z^{(0)}_M&=&\left(\frac{2^{\frac{d}{2}}}{8\pi(8\pi^2\alpha')^{\frac{d}{2}}}\right)^{\frac{1}{d}}
\int_0^{\infty}ds[2(d-2)+O(e^{-2s})]. \ee

Therefore \footnote{The limit $a\rightarrow0$ here is
corresponding to the limit $s\rightarrow\infty$ in \cite{book2}.}
\be
\frac{Z'_P}{Z'_M}=\lim_{s\rightarrow\infty}\frac{\left(\frac{1}{8\pi(8\pi^2\alpha')^{\frac{d}{2}}}\right)^{\frac{1}{d}}
\int_0^{\infty}ds[2(d-2)+O(e^{-2s})]}{\left(\frac{2^{\frac{d}{2}}}{8\pi(8\pi^2\alpha')^{\frac{d}{2}}}\right)^{\frac{1}{d}}
\int_0^{\infty}ds[2(d-2)+O(e^{-2s})]}=\frac{1}{\sqrt{2}} \ee

Then \be Z_{P+M}(y)&=&(2^{n})^2\int_0^1\frac{da^2}{2a^2}
\prod_{i=0}^pZ'_PZ_P(y_i,a^2)+2^n\int_{-1}^0\frac{da^2}{2a^2}
\prod_{i=0}^p\sqrt{2}Z'_PZ_P(y_i,a^2). \ee

It is convenient to regularize the  volume divergence of the
remaining coordinates as in \cite{0009148,0010108} by periodic
identification \be X^\mu\sim X^\mu+R^\mu,\mu=p+1,\cdots,d-1.\ee

To determine the correct normalization of the $X$ zero mode, we
use the same method as in \cite{0010108}. We find that the
normalization of the $X$ zero mode is $1/\sqrt{\pi}$ for the
M\"obius strip
\be
\int\frac{dX}{\sqrt{\pi}}\exp\left(-\frac{y}{8\pi}\int_0^{2\pi}d\theta
X^2\right)=\sqrt{2}\cdot\sqrt{\frac{2}{y}},\ee instead of
$1/\sqrt{2\pi}$ for the annulus

\be
\int\frac{dX}{\sqrt{2\pi}}\exp\left(-\frac{y}{8\pi}\int_0^{2\pi}d\theta
X^2\right)=\sqrt{\frac{2}{y}}.\ee

The string field action evaluated for the tachyon field is then
given by \be S^{P+M}_{\rm
1-loop}(y)&=&(2^{n})^2\int_0^1\frac{da^2}{2a^2}
\prod_{i=0}^pZ'_PZ_P(y_i,a^2)
\prod_{\mu=p+1}^{d-1}\left(\frac{R^\mu}{\sqrt{2\pi}}\right)\no\\
& & +2^n\int_{-1}^0\frac{da^2}{2a^2}
\prod_{i=0}^p\sqrt{2}Z'_PZ_P(y_i,a^2)
\prod_{\mu=p+1}^{d-1}\left(\frac{R^\mu}{\sqrt{\pi}}\right)\no\\
&=&(2^{n})^2\int_0^1\frac{da^2}{2a^2}
\prod_{i=0}^pZ'_PZ_P(y_i,a^2)
\prod_{\mu=p+1}^{d-1}\left(\frac{R^\mu}{\sqrt{2\pi}}\right)\no\\
& &2^{d/2}\cdot2^n\int_{-1}^0\frac{da^2}{2a^2}
\prod_{i=0}^pZ'_PZ_P(y_i,a^2)
\prod_{\mu=p+1}^{d-1}\left(\frac{R^\mu}{\sqrt{2\pi}}\right).\ee

For $d=10$ and $n=5$, which corresponding to $SO(2^5)=SO(32)$
gauge group, it can be written as \be\label{all} S^{P+M}_{\rm
1-loop}(y)=(32)^2\int_{-1}^1\frac{da^2}{2a^2}
\prod_{i=0}^pZ'_PZ_P(y_i,a^2)
\prod_{\mu=p+1}^{d-1}\left(\frac{R^\mu}{\sqrt{2\pi}}\right). \ee

Now we see that the divergence due to the integral over modulus
$a$ is cancelled if we consider the integration in (\ref{all}) as
the principal value integral. After performing the integration
over $a$, We can write down (\ref{all}) in term of tachyon fields
up to the first two orders \be S^{P+M}_{\rm 1-loop}(y)\simeq
C_1T_p\int
d^{10}X\,e^{-\frac{1}{2}T^2}\left[C_2(2\ln{2})\partial_\mu
T(X)\partial^\mu T(X)+1\right],\ee where $C_1$ and $C_2$ are
constants and $T_p$ is the $Dp$-brane tension.

Now we are ready to evaluate the one-loop effective action of
non-orientable open string field theory by adding up the planar,
unoriented and oriented non-planar diagrams \cite{0104099}, \be
S^{P+M+T}_{\rm 1-loop}&\simeq &C_1T_p\int
d^{10}X\,e^{-\frac{1}{2}T^2}\left[C_2(2\ln{2})\partial_\mu
T\partial T^\mu +1\right]\no\\
& &+T_p(\Lambda+C_3)\int d^{10}X\,e^{-\frac{1}{2}T^2}[(2\ln {2}
)\partial_\mu T\partial^\mu T+1]\no\\
&=&T_p\int d^{10}X\,e^{-\frac{1}{2}T^2}\left[(\Lambda+\Lambda_{\rm
finite})(2\ln{2})\partial_\mu T\partial^\mu
T+(\Lambda+\Lambda'_{\rm finite})\right],\ee where $\Lambda$ is
the cut-off, $\Lambda_{\rm finite}$ and $\Lambda'_{\rm finite}$
are constants.

Then the effective tachyon action, up to one-loop level, is
\be S&=&S^{\rm Disk}_{\rm tree}+S^{P+M+T}_{\rm 1-loop}\no\\
&\simeq &\frac{T_p}{g}\int
d^{10}X\,e^{-\frac{1}{4}T^2}\left[(2\ln{2})\partial_\mu
T\partial^\mu T+1\right]\no\\
& &T_p\int d^{10}X\,e^{-\frac{1}{2}T^2}\left[(\Lambda+\Lambda_{\rm
finite})(2\ln{2})\partial_\mu T\partial^\mu
T+(\Lambda+\Lambda'_{\rm finite})\right]\no\\
&=&\frac{T'_p}{\lambda'}\int
d^{10}X\,e^{-\frac{1}{2}T^2}\left[(2\ln{2})\partial_\mu
T\partial^\mu T+\left(\frac{1+\lambda'\Lambda'_{\rm
finite}}{1+\lambda'\Lambda_{\rm finite}}\right)\right],\ee where
$\lambda'$ is the renormalized effective string coupling which is
defined by \be \frac{1}{\lambda'}=\frac{1}{\lambda}+\Lambda,\ee
and \be \lambda=e^{-\frac{1}{4}T^2}.\ee $T'_p$ is the renormalized
$Dp$-brane tension which is defined by \be
T'_p=(1+\lambda'\Lambda_{\rm finite})T_p.\ee

\section{Conclusions}
\setcounter{footnote}{0}

In this work, we have studied type I superstring field theory,
which is a non-orientable open string field theory, with the
tachyonic field interaction on the boundary by using the BSFT
method. We obtained the tachyon effective action up to one-loop
level with the renormalized $Dp$-brane tension and renormalized
string coupling. The expected cancellation of divergence between
the planar graph and the unoriented graph was obtained. We noted
that this cancellation is due to the exact relationship between
the two integration constants $Z'_P$ and $Z'_M$, namely the
normalization. There is another method to fix the normalization,
by calculating the three tachyon amplitude and comparing it to the
same amplitude obtained by the cubic open string field theory.
This has been done for both bosonic string field theory
\cite{0009191} and superstring field theory \cite{0106231} on the
disk, but not at the one-loop level. An alternative way to obtain
the relationship between $Z'_P$ and $Z'_M$ is to directly
calculate the partition function by using the boundary states
method. We hope that these calculation will confirm our results.

\section*{Acknowledgements}

We thank Taijin Lee, P. Matlock and R. Rachkov for useful
discussion. This work is supported by a grant from Natural
Sciences and Engineering Research Council of Canada.

\appendix
\renewcommand{\theequation}{\thesection.\arabic{equation}}

\section{Derivation of Green's functions and the partition function}
\setcounter{equation}{0}

\subsection{Bosonic and Fermionic Green's functions}
In the bosonic case, to solve \be
\partial_z\bar\partial_{\bar z}G(z,w)=-2\pi\delta^{(2)}(z-w),
\ee with the boundary conditon \be \label{abc} (z\partial +\bar
z\bar\partial
-y_a)G_B(z,w)|_{\rho=a}&=&0,\no\\
G_B(z,w)|_{\rho=b}-G_B(-z,w)|_{\rho=b}&=&0, \ee we start with the
ansatz, \be\label{ansatz}
G_B(z,w)&=&-\ln |z-w|^2+C_1\ln |z|^2\ln |w|^2+C_2(\ln |z|^2+\ln |w|^2)+C_3\no\\
& &+\sum_{-\infty}^{\infty}a_k[(z\bar w)^k+(\bar z w)^k]
+\sum_{-\infty}^{\infty}b_k\left[\left(\frac{z}{w}\right)^k
+\left(\frac{\bar z}{\bar w}\right)^k\right] \ee

Inserting this ansatz into the boundary conditions (\ref{abc}),
expanding it by series and following the procedure as in
\cite{0101207,0104099}, we get \be & &G_B(z,w)\no\\ &=&-\ln
|z-w|^2+\ln |z|^2+\ln |w|^2+\frac{2-y\ln a^2}{y}\no\\ &
&-\sum_{n=1}^{\infty}\ln
\left[\left|1-(-1)^n\left(\frac{a}{b}\right)^{2n}\frac{z\bar
w}{a^2}\right|^2
\cdot\left|1-(-1)^n\left(\frac{a}{b}\right)^{2n}\frac{b^2}{z\bar
w}\right|^2\right.\no\\ &
&\hspace{1.5cm}\left.\cdot\left|1-(-1)^n\left(\frac{a}{b}\right)^{2n}\frac{z}{w}\right|^2
\cdot\left|1-(-1)^n\left(\frac{a}{b}\right)^{2n}\frac{w}{z}\right|^2\right]\no\\
& &-2\sum_{k=1}^{\infty}\frac{ya^{4k}}
{k(b^{2k}-(-1)^ka^{2k})[(k+y)b^{2k}-(-1)^k(k-y)a^{2k}]}
\left[\left(\frac{z\bar{w}}{a^2}\right)^k+\left(\frac{\bar z
w}{a^2}\right)^k\right]\no\\ &
&-2\sum_{k=1}^{\infty}\frac{yb^{2k}a^{2k}}
{k(b^{2k}-(-1)^ka^{2k})[(k+y)b^{2k}-(-1)^k(k-y)a^{2k}]}
\left[\left(\frac{b^2}{z\bar w}\right)^k+\left(\frac{b^2}{\bar z
w}\right)^k\right]\no\\ &
&-2\sum_{k=1}^{\infty}\frac{(-1)^kya^{2k}b^{2k}}
{k(b^{2k}-(-1)^ka^{2k})[(k+y)b^{2k}-(-1)^k(k-y)a^{2k}]}
\left[\left(\frac{z}{w}\right)^k+\left(\frac{\bar z}{\bar
w}\right)^k\right]\no\\ &
&-2\sum_{k=1}^{\infty}\frac{(-1)^kya^{2k}b^{2k}}
{k(b^{2k}-(-1)^ka^{2k})[(k+y)b^{2k}-(-1)^k(k-y)a^{2k}]}
\left[\left(\frac{w}{z}\right)^k+\left(\frac{\bar w}{\bar
z}\right)^k\right] \ee

In the Fermionic case, to solve \be
\bar{\partial}G_F(z,w)&=&-i\sqrt{zw}\delta^{(2)} (z-w),\no\\
\partial\tilde{G}_F(\bar{z},\bar{w})&=&+i\sqrt{\bar{z}\bar{w}}\delta^{(2)} (\bar{z}-\bar{w})
\ee with the boundary conditions \be
\left.\left(1-iy_a\frac{1}{\partial_\theta}\right)G_F\right|_{\rho=a}
&=&\left.\left(1+iy_a\frac{1}{\partial_\theta}\right)\tilde{G}_F\right|_{\rho=a},\no\\
G_F(z,w)|_{\rho=b}&=&\tilde{G}_F(-\bar z,\bar w)|_{\rho=b}. \ee we
start with the ansatz, \be iG_F(z,w)&=&\frac{\sqrt{zw}}{z-w}
+\sum_{r\in Z+\frac{1}{2}}a_r (z\bar{w})^r +\sum_{r\in
Z+\frac{1}{2}}a'_r
\left(\frac{z}{w}\right)^r,\no\\
i\tilde{G}_F(\bar{z},\bar{w})&=&-\frac{\sqrt{\bar{z}\bar{w}}}{\bar{z}-\bar{w}}
-\sum_{r\in Z+\frac{1}{2}}b_r (\bar{z}w)^r -\sum_{r\in
Z+\frac{1}{2}}b'_r \left(\frac{\bar{z}}{\bar{w}}\right)^r. \ee

After a straightforward, albeit lengthy, calculation we get the
following results for $G_F(z,w)$ and $\tilde{G}_F(\bar z,\bar w)$
\be iG_F(z,w) &=&\frac{\sqrt{zw}}{z-w}\no\\ &
&-\sum_{n=1}^{\infty}\frac{\sqrt{(-1)^n\left(\frac{a}{b}\right)^{2n}\frac{z\bar
w}{a^2}}} {1-(-1)^n\left(\frac{a}{b}\right)^{2n}\frac{z\bar
w}{a^2}}
+\sum_{n=1}^{\infty}\frac{\sqrt{(-1)^n\left(\frac{a}{b}\right)^{2n}\frac{b^2}{z\bar
w}}} {1-(-1)^n\left(\frac{a}{b}\right)^{2n}\frac{b^2}{z\bar
w}}\no\\ &
&-\sum_{n=1}^{\infty}\frac{\sqrt{(-1)^n\left(\frac{a}{b}\right)^{2n}\frac{z}{w}}}
{1-(-1)^n\left(\frac{a}{b}\right)^{2n}\frac{z}{w}}
+\sum_{n=1}^{\infty}\frac{\sqrt{(-1)^n\left(\frac{a}{b}\right)^{2n}\frac{w}{z}}}
{1-(-1)^n\left(\frac{a}{b}\right)^{2n}\frac{w}{z}}\no\\ &
&+2\sum_{r\geq\frac{1}{2}}\frac{ya^{4r}}
{(b^{2r}-(-1)^ra^{2r})[(r+y)b^{2r}-(r-y)(-1)^ra^{2r}]}
\left(\frac{z\bar{w}}{a^2}\right)^r\no\\ &
&-2\sum_{r\geq\frac{1}{2}}\frac{ya^{2r}b^{2r}}
{(b^{2r}-(-1)^ra^{2r})[(r+y)b^{2r}-(r-y)(-1)^ra^{2r}]}
\left(\frac{b^2}{z\bar{w}}\right)^r\no\\ &
&+2\sum_{r\geq\frac{1}{2}}\frac{(-1)^rya^{2r}b^{2r}}
{(b^{2r}-(-1)^ra^{2r})[(r+y)b^{2r}-(r-y)(-1)^ra^{2r}]}
\left(\frac{z}{w}\right)^r\no\\ &
&-2\sum_{r\geq\frac{1}{2}}\frac{(-1)^rya^{2r}b^{2r}}
{(b^{2r}-(-1)^ra^{2r})[(r+y)b^{2r}-(r-y)(-1)^ra^{2r}]}
\left(\frac{w}{z}\right)^r, \ee and \be
i\tilde{G}_F(z,w)&=&-\frac{\sqrt{\bar z\bar w}}{\bar z-\bar
w}\no\\ &
&+\sum_{n=1}^{\infty}\frac{\sqrt{(-1)^n\left(\frac{a}{b}\right)^{2n}\frac{\bar
z w}{a^2}}} {1-(-1)^n\left(\frac{a}{b}\right)^{2n}\frac{\bar z
w}{a^2}}
-\sum_{n=1}^{\infty}\frac{\sqrt{(-1)^n\left(\frac{a}{b}\right)^{2n}\frac{b^2}{\bar
z w}}} {1-(-1)^n\left(\frac{a}{b}\right)^{2n}\frac{b^2}{\bar z
w}}\no\\ &
&+\sum_{n=1}^{\infty}\frac{\sqrt{(-1)^n\left(\frac{a}{b}\right)^{2n}\frac{\bar
z}{\bar w}}} {1-(-1)^n\left(\frac{a}{b}\right)^{2n}\frac{\bar
z}{\bar w}}
-\sum_{n=1}^{\infty}\frac{\sqrt{(-1)^n\left(\frac{a}{b}\right)^{2n}\frac{\bar
w}{\bar z}}} {1-(-1)^n\left(\frac{a}{b}\right)^{2n}\frac{\bar
w}{\bar z}}\no\\ & &-2\sum_{r\geq\frac{1}{2}}\frac{ya^{4r}}
{(b^{2r}-(-1)^ra^{2r})[(r+y)b^{2r}-(r-y)(-1)^ra^{2r}]}
\left(\frac{\bar z w}{a^2}\right)^r\no\\ & &
+2\sum_{r\geq\frac{1}{2}}\frac{ya^{2r}b^{2r}}
{(b^{2r}-(-1)^ra^{2r})[(r+y)b^{2r}-(r-y)(-1)^ra^{2r}]}
\left(\frac{b^2}{\bar z w}\right)^r\no\\ &
&-2\sum_{r\geq\frac{1}{2}}\frac{(-1)^rya^{2r}b^{2r}}
{(b^{2r}-(-1)^ra^{2r})[(r+y)b^{2r}-(r-y)(-1)^ra^{2r}]}
\left(\frac{\bar z}{\bar w}\right)^r\no\\ &
&+2\sum_{r\geq\frac{1}{2}}\frac{(-1)^rya^{2r}b^{2r}}
{(b^{2r}-(-1)^ra^{2r})[(r+y)b^{2r}-(r-y)(-1)^ra^{2r}]}
\left(\frac{\bar w}{\bar z}\right)^r. \ee

\subsection{the Partition function}
To evaluate the partition function, we need to calculate the
propagators on the boundary, for both the bosonic and the
fermionic fields. At the boundary $\rho=a$, let $z=ae^{i\theta}$
and $w=ae^{i(\theta+\epsilon)}$, we have \be &
&\lim_{\epsilon\rightarrow 0}G_B(y,\epsilon;a,b)\no\\
&\equiv&\lim_{\epsilon\rightarrow
0}G_B(ae^{i\theta},ae^{i(\theta+\epsilon)})\no\\ &=&-2\ln
(1-e^{i\epsilon})-2\ln
(1-e^{-i\epsilon})+\frac{2}{y}-8\sum_{n=1}^{\infty}\ln
\left[1-(-1)^n\left(\frac{a}{b}\right)^{2n}\right]\no\\ &
&-4\sum_{k=1}^{\infty}\frac{ya^{4k}}
{k(b^{2k}-(-1)^ka^{2k})[(k+y)b^{2k}-(k-y)(-1)^ka^{2k}]}\no\\ &
&-4\sum_{k=1}^{\infty}\frac{yb^{4k}}
{k(b^{2k}-(-1)^ka^{2k})[(k+y)b^{2k}-(k-y)(-1)^ka^{2k}]}\no\\ &
&-8\sum_{k=1}^{\infty}\frac{(-1)^kya^{2k}b^{2k}}
{k(b^{2k}-(-1)^ka^{2k})[(k+y)b^{2k}-(k-y)(-1)^ka^{2k}]}\no\\
&\equiv&-2\ln (1-e^{i\epsilon})-2\ln (1-e^{-i\epsilon})+F(y,-a^2/b^2),
\ee where $F(y,-a^2/b^2)\equiv F(y,\epsilon=0,-a^2/b^2)$ are the terms which
are not singular when we take the limit $\epsilon=0$.

We define \be
G'_F(\epsilon,y)&\equiv&G'_F(ae^{+i\theta},ae^{+i(\theta+\epsilon)})
=\langle\psi(\theta)\frac{1}{\partial_\theta}\psi(\theta+\epsilon)\rangle
,\no\\
\tilde{G}'_F(\epsilon,y)&\equiv&\tilde{G}'_F(ae^{-i\theta},ae^{-i(\theta+\epsilon)})
=\langle\tilde{\psi}(\theta)\frac{1}{\partial_\theta}\tilde{\psi}(\theta+\epsilon)\rangle
. \ee

We expand $G'_F(\epsilon,y)$ and $\tilde{G}'_F(\epsilon,y)$, then
compare them to the expansion of $G_B(y,\epsilon,a^2/b^2)$. By a
straightforward calculation it can be shown that \be &
&G'_F(\epsilon,y)+\tilde{G}'_F(\epsilon,y)\no\\
&=&G_B(y,\epsilon,-a^2/b^2)-2G_B(2y,\frac{\epsilon}{2},\sqrt{-a^2/b^2}).
\ee

Now we are ready to calculate the partition function at the
boundary $\rho=a$. \be \frac{d}{dy}\ln
Z_M&=&-\frac{1}{8\pi}\int_0^{2\pi}d\theta\langle X^2
+\psi\frac{1}{\partial_\theta}\psi
+\tilde{\psi}\frac{1}{\partial_\theta}\tilde{\psi}\rangle\no\\
&\equiv&\lim_{\epsilon\rightarrow 0}-\frac{1}{8\pi
}\int_0^{2\pi}d\theta\langle X(\theta)X(\theta+\epsilon)
+\psi(\theta)\frac{1}{\partial_\theta}\psi(\theta+\epsilon)
+\tilde{\psi}(\theta)\frac{1}{\partial_\theta}\tilde{\psi}(\theta+\epsilon)\rangle\no\\
&=&\lim_{\epsilon\rightarrow 0}[G_B(y_a,y_b,\epsilon,-a^2/b^2)
+G'_F(\epsilon,y)+\tilde{G}'_F(\epsilon,y)]\no\\
&=&2\lim_{\epsilon\rightarrow 0}[G_B(y,\epsilon,-a^2/b^2)
-G_B(2y,\frac{\epsilon}{2},\sqrt{-a^2/b^2})]\no\\ &=&-8\ln
2+2F(y,-a^2/b^2)-2F(2y,\sqrt{-a^2/b^2}). \ee

Integrating over $y$, up to an integration constant, we get \be
\ln Z_M(a/b)&=&(2\ln 2)y -\frac{1}{2}\ln y\no\\ &
&+\sum_{k=1}^\infty \left\{\ln
\left[\left(1+\frac{2y_a}{k}\right)\left(1+\frac{2y_b}{k}\right)
-\left(1-\frac{2y_a}{k}\right)\left(1-\frac{2y_b}{k}\right)
\left(\sqrt{-\frac{a^2}{b^2}}\right)^k\right]\right.\no\\ &
&\left.-2\ln
\left[\left(1+\frac{y_a}{k}\right)\left(1+\frac{y_b}{k}\right)-\left(1-\frac{y_a}{k}\right)
\left(1-\frac{y_b}{k}\right)
\left(-\frac{a^2}{b^2}\right)^{k}\right]\right\}. \ee

Comparing this Green function to the one on an annulus, we find
that the only differences consist of replacing
$\left(\frac{a^2}{b^2}\right)$ by $\left(-\frac{a^2}{b^2}\right)$
and setting $y_b=0$.

The partition function, then, can be obtained as \be
Z_M(a/b)=Z'4^y\frac{Z^2_B(y,-a^2/b^2)}{Z_B(2y,\sqrt{-a^2/b^2})}, \ee
where $Z'$ is the integration constant, and \be
Z_B(y;a,b)=\frac{1}{\sqrt{y}}\cdot\prod_{k=1}^\infty\left[\left(1+\frac{y}{k}\right)
-\left(1-\frac{y}{k}\right)\left(-\frac{a^2}{b^2}\right)^{k}\right]^{-1}
\ee is the bosonic partition function on the M\"obius strip.

\subsection{Chan-Paton Factor}

Including the Chan-Paton factors, the boundary action can be
generalized  to be \be \label{nonab}e^{-S_{\rm bndy}}={\rm Tr} P
\exp\left\{-\frac{1}{8\pi}\int
d\theta\left[T^2(X)+(\psi^{\mu}\partial_{\mu}T)
\frac{1}{\partial_{\theta}}(\psi^{\nu}\partial_{\nu}T)\right]\right\}
.\ee

Substituting the tachyon profile \be
T(X)=\sum_{i=0}^{p}\sum_{\alpha=1}^{2n}u^\alpha_iX^i\gamma^\alpha
\ee in to (\ref{nonab}) and using the symmetry of $X_i^\alpha
X_j^\beta$ and the Clifford relatiions
$\{\gamma^\alpha,\gamma^\beta\}=2\delta^{\alpha\beta}$, we find
that the boundary terms are proportional to the identity matrix
\be e^{-S_{bndy}}&=&\mbox{Tr}P\exp\left\{-\frac{y_{ij}}{8\pi}\int
d\theta \left[X_i X_j+\psi_i\frac{1}{\partial\theta}
\psi_j\right]\cdot
\textbf{1}_{2^n\times 2^n}\right\}\no\\
&=&2^n\exp\left\{-\frac{y_{ij}}{8\pi}\int d\theta \left[X_i
X_j+\psi_i\frac{1}{\partial\theta} \psi_j\right]\right\}, \ee
where $y_{ij}\equiv\sum_{\alpha=1}^{2n}u^{\alpha}_iu^{\alpha}_j$.

Thus the boundary action decouples to the abelian ones. It is easy
to obtain the partition function as before.

\end{document}